\documentclass[prb,aps,twocolumn]{revtex4}

\usepackage{graphicx}
\usepackage{epsfig}
\usepackage{amsmath}
\usepackage{amssymb}
\usepackage{epstopdf}

\def\sigmaxc{\stackrel{\leftrightarrow}{\sigma}}

\begin{document}

\title{Stochastic time-dependent current-density functional theory:
a functional theory of open quantum systems}

\author{Roberto D'Agosta}
\email{dagosta@physics.ucsd.edu} \affiliation{Department of Physics,
University of California - San Diego, La Jolla, 92093, California}

\author{Massimiliano Di Ventra}
\email{diventra@physics.ucsd.edu} \affiliation{Department of
Physics, University of California - San Diego, La Jolla, 92093,
California}

\date{\today}

\begin{abstract}
The dynamics of a many-body system coupled to an external
environment represents a fundamentally important problem. To this
class of open quantum systems pertains the study of energy transport
and dissipation, dephasing, quantum measurement and quantum
information theory, phase transitions driven by dissipative effects,
etc. Here, we discuss in detail an extension of time-dependent
current-density-functional theory (TDCDFT), we named stochastic
TDCDFT [Phys. Rev. Lett. {\bf 98}, 226403 (2007)], that allows the
description of such problems from a microscopic point of view. We
discuss the assumptions of the theory, its relation to a density
matrix formalism, and the limitations of the latter in the present
context. In addition, we describe a numerically convenient way to
solve the corresponding equations of motion, and apply this theory
to the dynamics of a 1D gas of excited bosons confined in a harmonic
potential and in contact with an external bath.
\end{abstract}

\maketitle

\section{Introduction}
Density functional theory (DFT)~\cite{Hohenberg1964,Kohn1965} has found
widespread application in different fields ranging from materials
science to biophysics. Its original formulation dealt with the
ground-state properties of many-particle systems, but since then it
has been extended to the time domain,\cite{Runge1984,Ghosh1988,Vignale1996}
giving access to relevant information about the non-equilibrium properties
of many-body systems.\cite{Marques} According to which variable is
employed as the basic physical quantity of interest, namely the
density or the current density, these dynamical extensions are named
time-dependent DFT (TDDFT)\cite{Runge1984} or time-dependent
current-DFT (TDCDFT).\cite{Ghosh1988,Vignale1996}
The successes of these theories are
impressive and are mainly due to their conceptual and practical
simplicity which allows the mapping of the original interacting
many-body problem into an effective single-particle problem. From a
computational point of view this represents a major simplification
compared to other, equally valid, but computationally more demanding
many-body techniques.

Nevertheless, one needs to recognize that in its present form DFT
can only deal with systems evolving under Hamiltonian dynamics. This
leaves out a large class of physical problems related to the
interaction of a quantum system with one or several external
environments, namely the study of the dynamics of open quantum
systems.\cite{vanKampen,Gardiner1983,Breuer2002} Examples of such
problems include energy transport driven by a bath (e.g.,
thermoelectric effects), decoherence, phase transitions driven by
dissipative effects, quantum information and quantum measurement
theory, etc. The study of these problems from a microscopic point of
view would give unprecedented insight into the dynamics of open
quantum systems.

The present authors have recently extended DFT to the study of the
dynamics of open quantum systems by proving that, given an initial
condition and a set of operators that describe the system-bath
interaction, there is a one-to-one correspondence between the
ensemble-averaged current density and the external vector
potential.\cite{Diventra2007} This theory has been named stochastic
time-dependent current-DFT (S-TDCDFT).\cite{Diventra2007} Its
starting point is a {\em stochastic} Schr\"odinger equation
(SSE)\cite{vanKampen} which describes the time-evolution of the
state vector in the presence of a set of baths, which introduce
stochasticity in the system dynamics at the Markov-approximation
level, or if the baths' operators depend locally on time, it
represents a form of non-Markovian dynamics, whereby the interaction
of the baths with the system changes in time, but it carries
information only at the time at which the state vector is evaluated,
and not on its past dynamics [see Eq.~(\ref{stochasticse})]. A
practical application of S-TDCDFT to the decay of excited He and its
connection with quantum measurement theory can be found in
Ref.~\onlinecite{Bushong2007}.

If the Hamiltonian of the system does not depend on microscopic
degrees of freedom, such as the density or the current density, the
SSE is the stochastic unraveling of a quantum master equation for
the density matrix.\cite{Ghirardi1990,vanKampen} One could thus
argue that an equation of motion for the many-body density matrix is
an equally valid starting point for a functional theory of open
quantum systems.\cite{Burke2005} Unfortunately, this is not the case
for several reasons. These are mainly related to the lack of a
closed equation of motion for the density matrix when the
Hamiltonian of the system depends on microscopic degrees of freedom,
and the possible lack of positivity of the density matrix when the
Hamiltonian and/or bath operators are time-dependent: the Kohn-Sham
(KS) Hamiltonian is, by construction, always time-dependent in
TDDFT. As we will discuss in this paper, these fundamental drawbacks
do not pertain to the solution of the SSE, making it a solid
starting point to develop a stochastic version of DFT.

The paper is organized as follows. In Sec.~\ref{basic} we introduce
the basic notation of stochastic processes and equations of motion.
In Sec.~\ref{STDCDFT} we discuss S-TDCDFT and in Sec.~\ref{densmatC}
we make a connection with a density-matrix approach, showing the
limitations of the latter in the present DFT context. In
Sec.~\ref{numeric} we describe numerically convenient ways to solve
the equations of motion of S-TDCDFT, and in Sec.~\ref{bose} we apply
this theory to the time evolution of a gas of excited bosons
confined in a harmonic potential, interacting at a mean-field level
and coupled to an external time-independent environment. We finally
report our conclusions and plans for future directions in
Sec.~\ref{conclude}.

\section{Basic notation}\label{basic}
Let us consider a quantum-mechanical system of $N$ interacting
particles of charge $e$ subject to an external deterministic
perturbation. The Hamiltonian of this system is
\begin{equation}
\hat H=\sum_{i=1}^N \frac{\left[\hat p_i +e A_{ext}(\hat
r_i,t)\right]^2}{2m}+\sum_{i\not = j}^N U_{int}(\hat r_i-\hat r_j)
\label{h}
\end{equation}
where $A_{ext}(r,t)$ is the external vector potential and
$U_{int}(r)$ describes the particle-particle interaction
potential. We work here in a gauge in which the scalar potential is set
to vanish identically.

Let us assume that this quantum-mechanical system is coupled, via
given many-body operators, to one or many external environments that
can exchange energy and momentum with the system. If we assume that
the dynamics of each environment is described by a series of
independent memory-less processes, the dynamics of the system is
governed by the stochastic Schr\"odinger equation\cite{vanKampen}
($\hbar=1$ throughout the paper)
\begin{equation}
\begin{split}
\partial_t |\Psi(t)\rangle=&-i \hat H |\Psi(t)\rangle -
\frac12\sum_{\alpha}
\hat U_\alpha(t) |\Psi(t)
\rangle\\
&+\sum_\alpha l_\alpha(t) \hat V_\alpha(t) |\Psi(t)\rangle
\label{stochasticse}
\end{split}
\end{equation}
where $\hat U_\alpha$ and $\hat V_\alpha$ describe the coupling of
the system with the $\alpha$-th environment. We will see below that,
if we impose that the state vector has an ensemble-averaged norm
equal to one, then $\hat U_\alpha =\hat V^{\dag}_\alpha\hat
V_\alpha$ [Eq.~(\ref{fluctuation-dissipation})], which provides an
intuitive interpretation of these two operators in terms of
dissipation and fluctuations, respectively.

One can {\em postulate} that such stochastic equation governs the
dynamics of our open quantum system,\cite{Ghirardi1990} or, if the
Hamiltonian is {\em not} stochastic (i.e., it does not depend on
microscopic degrees of freedom such as the density or current
density), the SSE (\ref{stochasticse}) can be justified {\it  a
posteriori} by proving that it gives the correct time evolution of
the many-particle density matrix, namely it is the unraveling of a
quantum master equation for the density matrix (see also
Sec.~\ref{densmatC}).\cite{vanKampen} Or better yet, one can derive
the SSE (\ref{stochasticse}) from first principles using, e.g., the
Feshbach projection-operator method to trace out (from the total
Hamiltonian: system plus environment(s) and their mutual
interaction) the degrees of freedom of the environment(s) with the
assumption that the energy levels of the latter form a dense
set.\cite{Gaspard1999} In this way, one can in fact derive an
equation of motion more general than the SSE (\ref{stochasticse})
which is valid also for environments that do not fulfill the
memory-less approximation. In the memory-less approximation that
equation of motion reduces to the SSE
(\ref{stochasticse}).\cite{Gaspard1999}

Here we do not restrict the theory to time-independent $\hat
U_\alpha$ and $\hat V_\alpha$ operators, but we assume that the
dynamics of these operators is not affected by the presence of the
quantum mechanical system, i.e., we neglect possible feedback of the
quantum mechanical system on the external environments. Moreover, we
assume that $\hat V_{\alpha}$ and $\hat U_{\alpha}$ admit a power
expansion in time at any time.\footnote{This assumption is needed in
the proof of the theorem of Stochastic-TDCDFT, see Ref.
\onlinecite{Diventra2007}.} For instance, a sudden switch of the
system-bath coupling cannot be treated in our formalism. Finally we
admit that $\hat U_\alpha$ and $\hat V_\alpha$ may vary in space. In
the following the time and spatial arguments of $\hat U_\alpha$ and
$\hat V_\alpha$ are suppressed to simplify the notation.

We choose $\{\hat U_\alpha\}$ to be Hermitian operators. Indeed, any
anti-hermitian part of the $\hat U_{\alpha}$ operators is
effectively an external non-dissipative potential that can be
included in the Hamiltonian, and then via a gauge transformation in
the vector potential. In Eq.~(\ref{stochasticse}), $\{l_\alpha(t)\}$
are a set of Markovian stochastic processes
\begin{eqnarray}
&&\overline{l_\alpha(t)}=0,\\
&&\overline{l_\alpha(t)l_\beta(t')}=\delta_{\alpha,\beta}\delta(t-t')
\label{markov}
\end{eqnarray}
where the symbol $\overline{\cdots}$ indicates the stochastic
average over an ensemble of identical systems evolving according to
the stochastic Schr\"odinger equation (\ref{stochasticse}).

\subsection{It\^o calculus}
Clearly, Eq.~(\ref{stochasticse}) does not follow the ``standard''
rules of calculus. Indeed, since $|\Psi(t)\rangle$ is a stochastic
function of time its time derivative is not defined at any instant
of time, namely, the stochastic terms, $l_\alpha(t)\hat V_\alpha$
and the Markov approximation, Eq.~(\ref{markov}), make this equation
non-tractable with the standard calculus
techniques.\cite{Gardiner1983} In particular, one has to assign a
meaning to quantities like
\begin{equation}
\int_0^t f(t') l_\alpha(t') dt'\equiv\int_0^t f(t') dW_\alpha(t')
\label{integral}
\end{equation}
where $f(t)$ is a test function and $W_\alpha(t)$ is a Wiener
process such that\cite{vanKampen}
\begin{equation}
W_\alpha(t)=\int_0^tl_\alpha(t') dt'.
\label{intdW}
\end{equation}

There are many different ways to assign a physical and mathematical
interpretation to Eq.~(\ref{integral}). In this paper we use the
It\^o calculus\cite{Gardiner1983}
\begin{equation}
\int_0^t f(t') dW_\alpha(t')=\lim_{Q\to\infty}\sum_{i=1}^{Q-1} f(t_i)[W_
\alpha(t_{i+1})-W_
\alpha(t_i)]
\label{ito}
\end{equation}
where $\{t_i\}$ is a series of time steps such that $t_1=0$ and
$t_Q=t$. For instance, another possible choice is (Stratonovich)
\begin{equation}
\begin{split}
\int_0^t f(t)dW_\alpha(t)=\lim_{Q\to\infty}\sum_{i=1}^{Q-1} \frac{f(t_i)
+f(t_{i+1})}{2}\\
\times[W_\alpha(t_{i+1})-W_\alpha(t_i)].
\label{strat}
\end{split}
\end{equation}
In standard calculus, one can prove that the r.h.s. of
Eqs.~(\ref{ito}) and (\ref{strat}) are identical. However, this is
not true if $W_\alpha$ describes a stochastic process:
Eqs.~(\ref{ito}) and (\ref{strat}) bear different physical
interpretations, and it is then not surprising that they do not
coincide.

The Wiener process $W_\alpha$ describes the dynamics of the
fluctuations due to the environment and defines the coupling between
these fluctuations and the system. In considering the cumulative
effect of these fluctuations on the system we have (at least) two
possible choices. On the one hand, we may assume that the only
knowledge [embodied by the function $f(t)$ in (\ref{ito}) and (\ref{strat})]
on the system we have access to is that at times {\em preceding} the
instant at which a fluctuation takes place, thus leading to
Eq.~(\ref{ito}). Alternatively, we can assume that the response of
the system is determined by its properties ``in between'' the states
before and after the fluctuation has occurred, and thus
Eq.~(\ref{strat}) follows. This second interpretation is correct
only if the fluctuations of the environment are ``regular'', i.e.,
if the r.h.s. of the Eq.~(\ref{markov}) is replaced by a regular
function of $t-t'$. We will however, restrict ourselves to the case
in which Eq.~(\ref{markov}) is valid. This has some mathematical
advantages, and it is always possible to transform the results from
one formalism to the other by a simple mapping.\cite{vanKampen}

\subsection{Stochastic Schr\"odinger equation}
Once we have defined the rules of integration with respect to the
Wiener process, the SSE (\ref{stochasticse}) has to be interpreted
as
\begin{equation}
d|\Psi\rangle=\left[-i\hat H\,dt -\frac12\sum_\alpha \hat U_\alpha
\,dt + \sum_\alpha\hat V_\alpha dW_\alpha\right]|\Psi\rangle  ,
\label{diffstose}
\end{equation}
that is as an infinitesimal difference equation.\footnote{In the following,
see Sec.~\ref{numeric}, we will derive the finite difference equation that is
satisfied by the wavefunction $\Psi$.}

It is important to bear in mind that if the It\^o approach is used,
few of the rules of the standard calculus have to be modified. The
most important and relevant for our following discussion is the rule
of product differentiation or {\it chain
rule}.~\cite{Higham2001,Gardiner1983} Indeed, we have that if $\Psi$
and $\Phi$ are two states evolving according to the
SSE~(\ref{stochasticse}), then
\begin{equation}
d(\Psi\Phi)=\Psi d\Phi+(d\Psi)\Phi+d\Psi d\Phi.
\label{calculus}
\end{equation}
When Eq.~(\ref{diffstose}) is used to express Eq.~(\ref{calculus})
in terms of the Hamiltonian, the following simple rules of calculus
must be kept in mind~\cite{Higham2001}
\begin{equation}
dtdt=0;\;\;~dtdW_\alpha=0;\;\;~dW_\alpha dW_\beta\equiv
\delta_{\alpha,\beta}dt. \label{basicdifferentials}
\end{equation}
These relations, that we assume here valid without further
discussion, can be proved exactly in the It\^o approach to
stochastic calculus.\cite{Higham2001,Gardiner1983} The first two
mean that terms of order higher than $dt$ are neglected [from
Eqs.~(\ref{markov}) and~(\ref{intdW}) we see that $dW_\alpha \sim
\sqrt{dt}$] while the third ensures that the different environments
act independently on the dynamics of the quantum-mechanical system.

Eqs.~(\ref{calculus}) and~(\ref{basicdifferentials}) will be used as
basic rules of calculus throughout this paper. To simplify the
notation, in the following we will consider only one environment.
The generalization to many independent environments is
straightforward.

Having set the mathematical rules, we can now derive the equations
of motion for the particle density and current density. These
equations of motion will be our starting point to develop stochastic
TDCDFT. By using It\^o formula Eq. (\ref{calculus}) we immediately
obtain the equation of motion for the many-particle density (this is
a function of $N$ coordinates, including spin)
\begin{widetext}
\begin{equation}
d(\Psi^*\Psi)=\left[ i(\Psi^*\hat H)\Psi-i\Psi^*(\hat H\Psi) -\Psi^*
\hat U\Psi+(\Psi^* \hat V^ \dagger)(\hat V\Psi)\right]
dt+\left[(\Psi^* \hat V)\Psi+\Psi^*(\hat V\Psi)\right]dW.
\label{manyparticle-density}
\end{equation}
\end{widetext}
By integrating over all degrees of freedom of all particles, and
taking the ensemble average of the result, we obtain the equation of
motion for the ensemble-averaged total norm, $\overline N$
\begin{equation}
\frac{d\overline N}{dt}=\overline{\langle \hat V^\dagger \hat V-\hat
U\rangle}, \label{totalparticle}
\end{equation}
where the symbol $\langle \hat A \rangle$ indicates the standard
quantum-mechanical expectation value of the operator $\hat A$.
From
Eq.~(\ref{totalparticle}) we immediately see that if we assume $\hat
V^\dagger \hat V=\hat U$ we obtain that the state vector has an
ensemble-averaged constant norm. In the following, we are then going
to assume that
\begin{equation}
\hat V^\dagger \hat V\equiv\hat U.
\label{fluctuation-dissipation}
\end{equation}
This relation is reminiscent of the ``fluctuations-dissipation
theorem" which relates the dissipation that drives the system
towards an equilibrium state [the terms $\frac12\sum_\alpha \hat
U_\alpha\,dt$ in Eq.~(\ref{diffstose})] with the fluctuations
induced by the external environment (the terms $\sum_\alpha\hat
V_\alpha dW_\alpha$ in the same equation) and which drive the system
out of equilibrium. Here, however, this relation is not limited to a
system close to equilibrium but it pertains also to systems far from
equilibrium.

Using Eq.~(\ref{fluctuation-dissipation}), Eqs.~(\ref{diffstose})
and (\ref{manyparticle-density}) simplify to (for one environment)
\begin{equation}
d|\Psi\rangle=\left[-i\hat H |\Psi\rangle -\frac12 \hat V^\dagger
\hat V |\Psi\rangle\right]dt + \hat V |\Psi \rangle dW,
\label{stochasticse-fin}
\end{equation}
and
\begin{equation}
\begin{split}
d(\Psi^*\Psi)=&\left[ i(\Psi^*\hat H)\Psi-i\Psi^*(\hat H\Psi)-\Psi^*
\hat V^\dagger \hat V\Psi\right.\\
&\left.+(\Psi^* \hat V^\dagger)\hat V\Psi\right] dt\\
&+\left[(\Psi^* \hat V)\Psi+\Psi^*(\hat V\Psi)\right]dW,
\end{split}
\label{many-density}
\end{equation}
respectively.
Starting from Eqs.~(\ref{stochasticse-fin}) and (\ref{many-density}) we
can obtain the equation of
motion for the expectation value of any observable $\hat A$
\begin{eqnarray}
d\langle \hat A\rangle&=&(d\langle \Psi|)\hat A|\Psi\rangle+\langle
\Psi|\hat A(d|\Psi\rangle)+ (d\langle \Psi|)\hat A(d|\Psi\rangle)\nonumber\\
&=& \left\langle i[\hat A,\hat H]-\frac12 \left( \hat V^\dagger \hat V
\hat A+\hat A \hat V^\dagger
\hat V-2 \hat V^\dagger \hat A\hat V\right)\right\rangle dt\nonumber\\
&&+\langle\hat V^\dagger \hat A+\hat A\hat V\rangle dW. \label{eq-A}
\end{eqnarray}
The equation of motion for the ensemble-averaged expectation value
is obtained immediately from (\ref{eq-A}),
\begin{eqnarray}
\partial_t\overline{\langle \hat A\rangle}&=
&i \overline{\langle  [\hat A,\hat H] \rangle}\nonumber\\
&&-\frac12 \left( \overline{\langle\hat V^\dagger \hat V \hat
A\rangle+ \langle\hat A \hat V^ \dagger \hat V\rangle-2 \langle\hat
V^\dagger \hat A\hat V\rangle}\right)
\label{ope-dynamics}\\
&=&i\left\langle\left[\overline{\hat A},\hat H\right]\right\rangle
\nonumber\\
&&-\frac12 \left( \langle\hat V^\dagger \hat V \overline{\hat
A}\rangle+ \langle\overline{\hat A} \hat V^\dagger \hat V\rangle-2
\langle\hat V^\dagger \overline{\hat A}\hat V\rangle\right),
\label{ope-dynamics-simple}
\end{eqnarray}
where we have used $\overline{dW}=0$.

\begin{figure}[t!]
\includegraphics[width=8.5cm,clip]{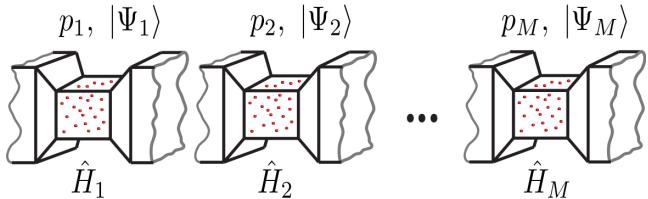}
\caption{(Color online) If the Hamiltonian depends on microscopic degrees of
freedom such as the particle or current density, it is different for
each element of the statistical ensemble represented by the
probabilities $p_i$ that the system is found in the state vector
$|\Psi_i\rangle $ of $M$ accessible states. A stochastic Hamiltonian
precludes the derivation of a closed equation of motion for the
density matrix [see discussion following
Eq.~(\ref{ope-dynamics-simple})].} \label{manyhamiltonians}
\end{figure}

In the last step we have also assumed that $\overline{\langle [\hat
A,\hat H] \rangle}= \langle [\overline{\hat A},\hat H]\rangle$. This
relation is valid only if $\hat H$ does {\it not} depend on any
stochastic field, i.e., it is not a {\em stochastic Hamiltonian}
which is different for the different elements of the statistical
ensemble (see Fig.~\ref{manyhamiltonians}). If, for example, the
particle-particle interaction in $\hat H$ is treated in the Hartree
approximation, then the last step in (\ref{ope-dynamics-simple}) is
not justified, and the equation of motion for the expectation value
of any operator $\hat A$ will not be given by
Eq.~(\ref{ope-dynamics-simple}) but by the more complex
Eq.~(\ref{ope-dynamics}).

\subsection{Quantum master equation}

For the simpler case in which the Hamiltonian is {\em not}
stochastic one can easily obtain a closed equation of motion for the
density matrix from the SSE. Quite generally we define
\begin{equation}
\hat \rho(t)\dot{=}\overline{|\Psi(t)\rangle\langle\Psi(t)|}\equiv
\sum_i\,p_i(t)\,|\Psi_i(t)\rangle
\langle\Psi_i(t)|,\label{SrhoTDCDFTA}
\end{equation}
where $|\Psi_i(t)\rangle$ is a pure state vector in the Hilbert
space of the system occurring in the ensemble with probability
$p_i(t)$, with $\sum_i\,p_i (t)=1$. Definition~(\ref{SrhoTDCDFTA})
is valid when the initial state of the system is pure. If the
initial state of the system is mixed with macro-state
$\{|\Psi_0^n\rangle,p^0_n\}$, then definition~(\ref{SrhoTDCDFTA}) of
statistical operator must include an extra summation
\begin{equation}
\hat \rho(t)\dot{=}\sum_n\,p^0_n\,\overline{|\Psi^n(t)\rangle \langle
\Psi^n(t)|},\label{SrhoTDCDFTAin}
\end{equation}
where $|\Psi^n(t) \rangle\equiv \{|\Psi^n_i(t)\rangle\}$ is the
ensemble of state vectors corresponding to the initial condition
$|\Psi_0^n\rangle$. Equation~(\ref{SrhoTDCDFTAin}) reduces
to~(\ref{SrhoTDCDFTA}) for a pure initial state
$\{p^0_n\}=\{1,0,\ldots,0\}$.

Using the definition~(\ref{SrhoTDCDFTAin}) of density matrix we can
define the ensemble average of any observable $\hat A$ as
\begin{equation}
\overline{\langle \hat A\rangle}=\textrm{Tr}\{ \hat \rho(t)\hat A\}.
\end{equation}

By using Eq.~(\ref{ope-dynamics-simple}) which is valid for any
observable, the many-particle density matrix operator follows the
equation of motion
\begin{equation}
\partial_t\hat \rho=-i \left[\hat \rho,\hat H\right] -\frac12
\left( \hat V^
\dagger \hat V \hat \rho+
\hat \rho \hat V^\dagger \hat V-2 \hat V \hat \rho\hat V^\dagger\right)
\label{density-matrix}
\end{equation}
which is the well-known quantum master equation (or Lindblad
equation if all operators, including the Hamiltonian, do not depend
on time).\cite{Lindblad1976,Gardiner1983,Breuer2002}

We stress once more that, in order to derive this quantum master
equation, we have assumed that the Hamiltonian does not depend on
any stochastic field. Otherwise, our starting point would have been
Eq.~(\ref{ope-dynamics}) and {\em no closed equation of motion} for
the density matrix could have been obtained.

Note that this is true even if the system does not interact with an
external environment but its state is mixed. A stochastic
Hamiltonian prevents us from writing a closed equation of motion for
the density matrix while the SSE~(\ref{stochasticse-fin}) contains
this case quite naturally: one simply evolves the system dynamics
over the ensemble of stochastic Hamiltonians and then averages the
resulting dynamics. This point is particularly relevant in DFT where
the KS Hamiltonian {\em does} depend on microscopic degrees of
freedom, and it is thus generally stochastic.~\cite{Diventra2007}

There is another important reason for not using the quantum master
equation~(\ref{density-matrix}) in a DFT approach. In fact, it is
only when the Hamiltonian of the system and the bath operators are
time-independent that one can prove that the density matrix solution
of Eq.~(\ref{density-matrix}) fulfills the usual requirements of a
``good'' statistical operator, i.e., that at any instance of time
its trace is conserved, the operator is hermitian and that it
remains a definite-positive operator, namely for any state $\Phi$ in
the Hilbert space
\begin{equation}
\langle\Phi|\hat \rho(t) |\Phi\rangle \geq 0. \label{positivity}
\end{equation}
The reason for these restrictions is because a dynamical semi-group
(in the exact mathematical sense) can only be defined for
time-independent
Hamiltonians.
\cite{Lindblad1976,Maniscalco2004,Maniscalco2004a,Whitney2008}

It is important to realize that an approach based on the
SSE~(\ref{stochasticse-fin}) does not suffer from this drawback: the
density matrix~(\ref{SrhoTDCDFTAin}) constructed from the SSE is
{\em by definition} positive at any time.

All of this points once more to the fact that
Eq.~(\ref{density-matrix}) is {\em not} a good starting point to
build a stochastic version of DFT. We will expand a bit more on
these issues in Sec.~\ref{densmatC}. In Sec.~\ref{bose} we will
provide an explicit example that shows that
Eq.~(\ref{density-matrix}) leads to the wrong dynamics in the
presence of interactions among particles.

\subsection{Continuity equation}
We can use the general result Eq.~(\ref{ope-dynamics}) to derive the
equation of motion for the ensemble-averaged particle density. Let
us define the ensemble-averaged density
\begin{equation}
\overline {n(r,t)}=\langle \overline{{\hat n}(r,t)}\rangle,
\end{equation}
and current density
\begin{equation}
\overline {j(r,t)}=\langle \overline{\hat j(r,t)}\rangle,
\end{equation}
where the current operator is defined as
\begin{equation}
\hat j(r,t)=\frac{1}{2}\sum_i\left\{\delta(r-\hat r_i), \hat
v_i\right\}
\end{equation}
with
\begin{equation}
\hat v_i=\frac{\hat p_i + e A_{ext}(\hat r_i,t)}m, \label{velocity}
\end{equation}
the velocity operator of particle $i$, and the symbol $\{\hat A,
\hat B\}\equiv (\hat A \hat B + \hat B \hat A)$ is the
anti-commutator of any two operators $\hat A$ and $\hat B$. From
Eq.~(\ref{ope-dynamics}) we then get
\begin{equation}
\begin{split}
\frac {\partial \overline{n(r,t)}}{\partial t}=&-\nabla \cdot\overline{j(r,t)}\\
&+\frac12
\overline{\left\langle 2 \hat V^\dagger \hat n(r,t) \hat V- \hat
V^\dagger
\hat V \hat n(r,t)-\hat
n(r,t) \hat V^\dagger \hat V \right\rangle}.
\label{eq-density}
\end{split}
\end{equation}
The last term on the right-hand side of Eq.~(\ref{eq-density}) is
identically zero for bath operators that are local in
space,~\cite{Frensley1990} namely
\begin{equation}
\overline{\left\langle 2 \hat V^\dagger \hat n(r,t) \hat V- \hat
V^\dagger \hat V \hat n(r,t)-\hat n(r,t) \hat V^\dagger \hat V
\right\rangle}\equiv 0. \label{bathcondition}
\end{equation}

Most transport theories satisfy this requirement since the action
that a true bath does on the system is derived from microscopic
mechanisms (e.g., inelastic processes) which are generally
local.~\cite{Frensley1990} If this were not the case, then this term
would represent {\em instantaneous} transfer of charge between
disconnected -- and possibly macroscopically far away -- regions of
the system without the need of mechanical motion, represented by the
first term on the right-hand side of Eq.~(\ref{eq-density}). This
instantaneous ``action at a distance'' is reminiscent of the
postulate of wave-packet reduction whereby the system may change its
state in a non-unitary way upon measurement.

Here, it is the result of the memory-less approximation that
underlies the stochastic Schr\"odinger
equation~(\ref{stochasticse-fin}). By assuming that the bath
correlation times are much shorter than the times associated with
the dynamics of the system (in fact, in the Markov approximation
these correlation times are assumed zero), we have lost information
on the microscopic interaction mechanisms at time scales on the
order of the correlation times of the bath. In other words, we have
coarse-grained the time evolution of our system, and we are
therefore unable to follow its dynamics on time scales smaller than
this time resolution.~\cite{Gebauer2004}

In the following, we will assume that the
condition~(\ref{bathcondition}) is identically satisfied or, if it
is not, at any given time, given a physical ensemble-averaged
current density, a unique solution for the ensemble-averaged density
can be found from Eq.~(\ref{eq-density}).

\subsection{Equation of motion for the current density}
Similarly, we can derive the equation of motion for the
ensemble-averaged
current density~\cite{Diventra2007}
\begin{eqnarray}
\partial_t \overline{j(
r,t)}&=&\frac{\overline {n(r,t)}}{m} \partial_t {A_{ext}}({
r},t) -\frac{\overline {{j}({r},t)} }{m}\times\left[\nabla \times
{A_{ext} }
({r},t)\right]\nonumber
\\
&&+\frac{\langle \overline{\hat {\mathcal F}({
r},t)}\rangle}{m} +\langle\overline{\hat {\mathcal G}({r},t)}\rangle
\label{currenteq}
\end{eqnarray}
where we have defined\footnote{In Eq.~(\ref{forces}), $\nabla_j$
contains the derivatives with respect to the coordinates of the
$j$-th particle, i.e., in 3D
$\nabla_j=(\partial_{x_j},\partial_{y_j},\partial_{z_j})$.}
\begin{equation}
\begin{split}
\label{forces}
\hat {\mathcal G}({ r},t)&=\hat V^\dagger \hat j({ r},t) \hat
V  -\frac12 \hat j({ r},t) \hat V^\dagger \hat V
-\frac12 \hat V^\dagger \hat V \hat j({ r},t),\\
\hat {\mathcal F}({ r},t)&=-\sum_{i\not = j}\delta({ r}-\hat r_i)
\nabla_j U\left(\hat r_i-\hat r_j\right)+m \nabla
\cdot \hat \sigmaxc({ r},t)
\end{split}
\end{equation}
with the stress tensor $\hat \sigmaxc({ r},t)$ given by
\begin{equation}
\hat \sigma_{i,j}({ r},t)=-\frac{1}{4}\sum_k \{\hat v_i,\{\hat
v_j,\delta({ r}-\hat r_k)\}\}.
\label{stresstensor}
\end{equation}

The first two terms on the rhs of Eq.~(\ref{currenteq}) describe the
effect of the applied electromagnetic field on the dynamics of the
many-particle system; the third is due to particle-particle
interactions while the last one is the ``force'' density exerted by
the bath on the system. This last term is responsible for the
momentum transfer between the quantum-mechanical system and the
environment.

\section{Stochastic Time-Dependent Current-Density Functional
Theory}\label{STDCDFT}

Having discussed the physical and mathematical requirements for the
problem we are interested in, we can now state the following theorem
of stochastic time-dependent current-DFT.~\cite{Diventra2007}

{\it Theorem:} Consider a many-particle system described by the
dynamics in Eq.~(\ref{stochasticse}) with the many-body Hamiltonian
given by Eq.~(\ref{h}). Let $\overline {n({ r},t)}$ and $\overline
{{ j}({ r},t)}$ be the ensemble-averaged single-particle density and
current density, respectively, with dynamics determined by the
external vector potential ${A_{ext}}({ r},t)$ and bath operators
$\{\hat V_{\alpha}\}$. Under reasonable physical assumptions, given
an initial condition $|\Psi_0\rangle$, and the bath operators
$\{\hat V_{\alpha}\}$, another external potential  $A'_{ext}({
r},t)$ which gives the same ensemble-averaged current density, must
necessarily coincide, up to a gauge transformation, with $A_{ext}({
r},t)$.

The details of the proof of this theorem can be found in Ref.~
\onlinecite{Diventra2007}. Here, we just mention that the initial
condition need not be a pure state for the theorem to be valid, but
may include also the case of mixed initial states.

The general idea of the proof, following similar ones proposed by
van Leeuwen\cite{vanLeeuwen1999} and Vignale\cite{Vignale2004}, is
to show that the external potential $A'_{ext}$ is completely
determined, via a power-series expansion in time, by $A_{ext}$, the
ensemble-averaged current density, the initial condition, and the
bath operators.

A lemma of the theorem states that any ensemble-averaged current
density that is interacting $A-$representable is also
non-interacting $A-$representable. (A current density is
$A-$representable if and only if it can be generated by the
application of an external potential $A$.) This implies that if an
ensemble-averaged current density can be generated in an interacting
system by a given vector potential, then it exists a non-interacting
system (the KS system) in which we can obtain the same current
density by applying {\em another} suitable vector potential, we will
call from now on $A_{eff}$.

This is opposed to the general result that an interacting
$V-$representable current density (namely one that is generated by a
scalar potential $V$) is not necessarily non-interacting
$V-$representable.~\cite{DAgosta2005a} In particular, it has been
shown that the mapping between the current density and the scalar
potential is not invertible.~\cite{DAgosta2005a} This result shows
that time-dependent DFT does not necessarily provide the exact
current density, even if the exact exchange-correlation potential is
known (albeit it provides the exact {\em total} current for a finite
and closed system~\cite{DiVentra2004a}). With some hindsight this is
not surprising since there is clearly no one-to-one correspondence
between a scalar and a vector.

\subsection{The stochastic Kohn-Sham equations}
Let us now assume that we know exactly the vector potential
$A_{eff}$ that generates the exact current density in the
non-interacting system. By construction, the system follows the
dynamics induced by the SSE (for a single bath operator)
\begin{equation}
d|\Psi_{KS}\rangle =\left(-i\hat H_{KS}-\frac12 \hat V^\dagger \hat
V\right)|\Psi_{KS}\rangle dt+\hat V |\Psi_{KS}\rangle dW
\label{ksse}
\end{equation}
where $|\Psi_{KS}\rangle$ is a Slater determinant of single-particle
wave-functions and
\begin{equation}
\hat H_{KS}=\sum_{i=1}^N \frac{\left[\hat
p_i+eA_{eff}(r_i,t)\right]^2}{2m} \label{ksh}
\end{equation}
is the Hamiltonian of non-interacting particles.

Note that for a general bath operator acting on many-body
wave-functions one cannot reduce Eq.~(\ref{ksse}) to a set of
independent single-particle equations. The reason is that our
theorem guarantees that one can decouple the quantum correlations
due to the direct interaction among particles, but one cannot
generally decouple the statistical correlations induced by the
presence of the environment. These affect the population of the
single-particle states of the quantum-mechanical system, while the
quantum correlations are taken into account to all orders by the
external potential $A_{eff}$ acting on the KS system. It is only
when the bath operators act on single-particles or on the density
that one can write Eq.~(\ref{ksse}) as a set of equations of motion,
one for every KS single-particle state.~\cite{Bushong2007}

\subsection{Initial conditions}
The initial condition for the time evolution of the KS system has to
be chosen such that the ensemble-averaged particle and current
densities coincide with those of the many-body interacting system.
Again, it is important to stress that in going from the interacting
system to its non-interacting doppelg\"anger, the bath operator is
{\it not} modified. On the other hand, the bath operator $\hat V$
generally induces transitions between many-body states of the
interacting Hamiltonian (\ref{h}). Therefore, when represented in
the non-interacting basis of the KS Hamiltonian it may connect many
different single-particle KS states. It has been argued that this
way the KS system will never reach a stationary state even if the
coupling with the environment is purely
dissipative.~\cite{Burke2005} It would be thus tempting to modify
the bath operator to force the KS system into an equilibrium with
the external environment.~\cite{Burke2005} This procedure, however,
breaks the theorem we have proved, and contains approximations of
unknown physical meaning.

In reality, if the true many-body system reaches equilibrium with
the environment, then the ensemble-average current and particle
density would attain a stationary limit. Since, these are the only
two physical quantities that the KS system needs to reproduce, the
question of whether the latter is in equilibrium with the
environment or not has no physical relevance.

\subsection{The exchange-correlation vector potential}
The vector potential $A_{eff}(r,t)$, acting on the KS system is
generally written as the sum of two contributions
\begin{equation}
A_{eff}(r,t)=A_{ext}(r,t) +A_{h-xc}(r,t),
\end{equation}
where $A_{ext}(r,t)$ is the vector potential applied to the {\em
true} many-body system, and $A_{h-xc}(r,t)$ is the vector potential
whose scope is to mimic the correct dynamics of the
ensemble-averaged current density. From the theorem we have proven,
$A_{h-xc}(r,t)$ is a functional of the average current density
$\overline{j(r,t')}$, for $t'\leq t$ (namely, it is
history-dependent), the initial condition {\it and} the bath
operator $\hat V$.~\cite{Diventra2007}

A common expression would isolate from $A_{h-xc}(r,t)$ the Hartree
interaction contribution from the ``rest'' due to the particle
exchange and correlation, namely one makes the {\em ansatz}
\begin{equation}
A_{h-xc}(r,t)=A_h(r,t)+A_{xc}(r,t)
\label{ahxc}
\end{equation}
where $A_h(r,t)$ is the Hartree contribution to the vector potential
($t_0$ is the initial time)
\begin{equation}
A_h(r,t)=\int_{t_0}^t dt'\nabla \int dr'
\frac{\overline{n(r',t')}}{|r-r'|}.\label{HartreeA}
\end{equation}
The other contribution, $A_{xc}(r,t)$ is again a functional of the
average current density $\overline{j(r,t')}$, for $t'\leq t$, the
initial condition, and the bath operator $\hat
V$,\cite{Diventra2007}
\begin{equation}
A_{xc}(r,t)=A_{xc}\left[\overline{j(r,t')}, |\Psi_0\rangle , \hat
V\right].
\end{equation}

In the present case, however, particular care needs to be applied to
the above {\em ansatz}. We have written the Hartree contribution in
terms of the ensemble-averaged density. This choice, however,
requires that the exchange-correlation vector potential included
also the statistical correlations of the direct Coulomb interaction
at different points in space. These correlations may be very large,
and possibly much larger than the Coulomb interaction between the
average densities. The ambiguity here, compared to the pure-state
case, is because in a mixed state, quite generally the
ensemble-average of the direct Coulomb interaction energy contains
statistical correlations between densities at different points in
space, namely
\begin{equation}
\int dr \int dr' \frac{\overline{\langle \hat n(r)\rangle\,\langle
\hat n(r')\rangle }}{|r-r'|}\neq \int dr \int dr' \frac{\langle
\overline{\hat n(r)}\rangle \; \langle \overline{\hat n(r')}\rangle
}{|r-r'|}.
\end{equation}
In actual calculations, one would instead use the form of the
Hartree potential in terms of the density {\em per element} of the
ensemble. This choice~(\ref{HartreeA}) makes the KS
Hamiltonian~(\ref{ksh}) stochastic, and therefore, as discussed in
Sec.~\ref{basic}, no closed equation of motion for the many-particle
KS density matrix can be obtained.

Finally, in view of the fact that one can derive a Markovian
dynamics only on the basis of a weak interaction with the
environment,\cite{vanKampen,Gardiner1983} as a first approximation,
one may neglect the dependence of the exchange-correlation vector
potential on the bath operator, and use the standard functionals of
TD-DFT and TD-CDFT.\cite{Marques,Vignale1996} [Like the Hartree
term, these functionals would also contribute to the stochasticity
of the KS Hamiltonian~(\ref{ksh}).] This seems quite reasonable, but
only comparison with experiments and the analysis of specific cases
can eventually support it. We thus believe that more work in this
direction will be necessary.

\section{Connection with a density-matrix approach}
\label{densmatC}

From the KS Slater determinants $|\Psi^i_{KS}\rangle$, solutions of
the KS equation~(\ref{ksse}) occurring with weight $p_i(t)$ ($\sum_i
p_i(t)=1$) in the ensemble, we can construct the {\em many-particle}
KS density matrix [from Eq.~(\ref{SrhoTDCDFTA})]
\begin{equation}
\hat
\rho_{KS}(t)=\overline{|\Psi^i_{KS}(t)\rangle\langle\Psi^i_{KS}(t)|}
\equiv
\sum_i\,p_i(t)\,|\Psi^i_{KS}(t)\rangle
\langle\Psi^i_{KS}(t)|.\label{SrhoTDCDFTAKS}
\end{equation}
This density matrix is, by construction, always positive-definite,
despite the fact that the KS Hamiltonian and possibly the $\hat V$
operator are time dependent. Since in general the bath operator acts
on many-particle states, this many-particle KS density matrix cannot
be reduced to a set of single-particle density matrices (see
Sec.~\ref{numeric} for a numerical {\em ansatz} suggested in
Ref.~\onlinecite{Pershin2008} to simplify the calculations).

We note first that, in principle, if we knew the exact functional
$A_{h-xc}$ as a functional of the averaged current density, then the
KS Hamiltonian~(\ref{ksh}) would not be stochastic and we could
derive the equation of motion of the many-particle KS density
matrix~(\ref{SrhoTDCDFTAKS}). This equation of motion would be
Eq.~(\ref{density-matrix}) with $\hat H$ replaced by $\hat H_{KS}$.

It is important to point out, however, that it is only when we start
from the stochastic KS equation~(\ref{ksse}) to construct the
density matrix~(\ref{SrhoTDCDFTAKS}) that we are guaranteed that the
solution of Eq.~(\ref{density-matrix}) for the KS density matrix
maintains positivity at any time. The reverse is not necessarily
true: the equation of motion~(\ref{density-matrix}) for the KS
density matrix may, for an arbitrary bath operator $\hat V$ or
initial conditions, provide non-physical solutions. In other words,
Eq.~(\ref{density-matrix}) admits more solutions than physically
allowed, while the SSE always provides a physical state of the
system dynamics.

We also stress once more that any approximation to $A_{h-xc}$, makes
the KS Hamiltonian stochastic, namely one Hamiltonian for each
element of the ensemble, thus making the density-matrix formalism of
limited value. In fact, by insisting on using
Eq.~(\ref{density-matrix}) with these approximations would amount to
introducing uncontrollable approximations in the system dynamics
which entail neglecting important statistical correlations induced
by the bath (see also discussion in Sec.~\ref{bose}).

To see this point explicitly, let us consider the equation of motion
for an arbitrary operator acting on the KS system that evolves
according to Eq.~(\ref{eq-A}) with $\hat H_{KS}$ replacing $\hat H$
\begin{equation}
\begin{split}
d\langle \hat A\rangle=&\left\langle i [\hat A,\hat H_{KS}] -\frac12
\left( \hat V^\dagger \hat V
\hat A+\hat A \hat V^\dagger \hat V-2 \hat V \hat A\hat V^\dagger\right)
\right\rangle dt\\
&+\langle\hat V^\dagger \hat A+\hat A\hat V\rangle dW.
\end{split}
\label{eq-Aks}
\end{equation}
Now we take the ensemble average of this equation in order to obtain
the equation of motion for the ensemble-averaged quantities.
However, since now $H_{KS}$ is a stochastic Hamiltonian, then the
ensemble average and the commutator between $\hat A$ and $\hat
H_{KS}$ {\it do not} commute, i.e., in general we expect that
\begin{equation}
\overline{[\hat A,\hat H_{KS}]}\not \equiv [\overline{\hat A},\hat
H_{KS}],
\end{equation}
which implies that
\begin{equation}
\begin{split}
\partial_t\hat \rho_{KS}\neq &-i \left[\hat \rho_{KS},\hat
H_{KS}\right]
\\
&-\frac12 \left( \hat V^\dagger \hat V \hat \rho_{KS}+\hat \rho_{KS}
\hat V^\dagger \hat V-2 \hat V \hat \rho_{KS}\hat V^\dagger\right),
\label{ks-lindblad}
\end{split}
\end{equation}
namely there is no closed equation of motion for the KS density
matrix.

In fact, the correct procedure is to evolve the system for {\em
every} realization of Hamiltonians, and then average over these
realizations, which is what a solution of the SSE~(\ref{ksse}) would
provide.

\section{Numerical solution of the SSE}\label{numeric}

\subsection{Finite difference equation}
We now discuss practical implementations of the SSE~(\ref{ksse}).
First of all we realize that, in going from the differential
equation~(\ref{ksse}) to a finite difference equation that can be
solved on a computer, one has to bear in mind that $dW$ is on the
order of $\sqrt{dt}$. Then one has to expand the equation for the
finite differences and keep terms of {\it second order} in $dW$ that
correspond to {\it first order} in $dt$.

Here we write down the correct finite difference equation starting
from~(\ref{ksse}). In the following we assume that the state vector
$\Psi_{KS}$ is a regular function of time, position, and the
Wiener process $W$, i.e., we assume that the derivatives
\begin{equation}
\frac{\partial \Psi_{KS}}{\partial W},~~\frac{\partial^2 \Psi_{KS}}
{\partial W
\partial W},~\ldots
\end{equation}
exist and are regular.

Let us define $dt=t-t'$ a small time interval over which we
integrate the equation of motion for $\Psi_{KS}$. If we expand in
series the increment $d\Psi_{KS}=\Psi_{KS}(t)-\Psi_{KS}(t')$ we
have,
\begin{eqnarray}
d\Psi_{KS}&=&\frac{\partial \Psi_{KS}}{\partial t} dt+\frac{\partial
\Psi_{KS}}{\partial W} dW+\frac{1}{2}
\frac{\partial^2 \Psi_{KS}}{\partial W\partial W} dW
dW+\ldots\nonumber\\
&=&\left(\frac{\partial \Psi_{KS}}{\partial t}+\frac{1}{2}
\frac{\partial^2 \Psi_{KS}}{\partial W\partial W}\right) dt+
\frac{\partial \Psi_{KS}}{\partial W}  dW.
\end{eqnarray}
A direct term-by--term comparison with Eq.~(\ref{ksse}) tells us
that there is a correspondence between $\partial /{\partial W}$ and
$\hat V$ so that a finite difference scheme can now be implemented
such that the equation of motion
\begin{eqnarray}
\Delta \Psi_{KS}&=&\left(\frac{\partial \Psi_{KS}}{\partial t}\right)
\Delta t+\left(\frac{\partial \Psi_{KS}}{\partial W}\right)\Delta
W\nonumber\\
&=&\left(-iH_{KS}-\frac{1}{2}V^\dagger V-\frac{1}{2}\hat V^2\right)
\Psi_{KS} \Delta t \nonumber\\
&&+\hat V\Psi_{KS} \Delta W,
\label{num-dynamics}
\end{eqnarray}
is the correct first-order equation in $\Delta t$.
Eq.~(\ref{num-dynamics}) can now be solved by a variety of different
numerical techniques, and we refer the reader to other work for a
discussion of such methods.\cite{Kloeden1997,Breuer2002,Wilkie2004,Wilkie2005}

The important point is that one evolves these equations in time for
every realization of the stochastic process and then averages over
the different realizations (in Sec.~\ref{bose} we give an explicit
example of such calculation showing the convergence of the results
with the number of realizations).

\subsection{The non-linear SSE}
The norm of the state vector solution of the SSE~(\ref{ksse}) is
preserved on average but not for every realization of the stochastic
process.\cite{vanKampen,Diventra2007} This may slow down the convergence
of the results as a function of the number of realizations of the
stochastic process. It is thus more convenient to solve a {\em
non-linear} SSE which gives an equivalent solution as the {\em
linear} SSE~(\ref{ksse}). This can be easily done by first
calculating the differential [in the It\^o sense~(\ref{ito})] of the
square modulus of $|\Psi_{KS}\rangle$
\begin{eqnarray}
d||\Psi_{KS}||^2&=&\langle
\Psi_{KS}|(\hat V^{\dag}+\hat V)|\Psi_{KS}\rangle dW\nonumber \\
&\equiv& 2 R\,||\Psi_{KS}||^2 dW,
\end{eqnarray}
where we have defined
\begin{equation}
R=\frac{1}{2}\frac{\langle \Psi_{KS}|(\hat V^{\dag}+\hat
V)|\Psi_{KS}\rangle}{||\Psi_{KS}||^2}.
\end{equation}
By using the power expansion
\begin{eqnarray}
d||\Psi_{KS}||&=&d\sqrt{||\Psi_{KS}||^2}\nonumber\\
&=&\frac{1}{2\sqrt{||\Psi_{KS}||^2}}d
||\Psi_{KS}||^2 \nonumber\\
&&-\frac{1}{8(\sqrt{||\Psi_{KS}||^2})^3}d ||\Psi_{KS}||^2
d||\Psi_{KS}||^2,
\end{eqnarray}
we can derive the equation of motion for the state vector normalized
at every realization of the stochastic process:
\begin{equation}
|\Phi_{KS}\rangle =\frac{|\Psi_{KS}\rangle}{||\Psi_{KS}||}
\end{equation}
which is (see also Ref.~\onlinecite{Ghirardi1990})
\begin{eqnarray}
d|\Phi_{KS}\rangle&=&\left [ -i\hat H_{KS} -\frac12 \hat
V^{\dag}\hat
V +R\hat V -\frac12 R^2\hat 1\right ]|\Phi_{KS}\rangle dt\nonumber\\
&& + (\hat V - R\hat 1)|\Phi_{KS}\rangle dW.
\end{eqnarray}
This non-linear equation of motion, by construction, is equivalent
to the linear SSE~(\ref{ksse}).

\begin{widetext}
The finite-difference equation for this case is
\begin{eqnarray}
\Delta \Phi_{KS}&=&\left(\frac{\partial \Phi_{KS}}{\partial t}\right)
\Delta t+\left(\frac{\partial \Phi_{KS}}{\partial W}\right)\Delta
W\nonumber\\
&=&\left(-iH_{KS}-\frac{1}{2}V^\dagger V-\frac{1}{2}\hat
V^2
+R\hat V -\frac12 R^2\hat 1\right)\Phi_{KS} \Delta t
+(\hat V - R\hat 1)\Phi_{KS} \Delta W.\label{num-dynamicsphi}
\end{eqnarray}
\end{widetext}

\subsection{Single-particle order-$N$ scheme}
Due to the presence of the environment, it is still a formidable
task to solve the equations of motion of S-TDCDFT for arbitrary bath
operators. In fact, as we have already discussed, the bath operators
generally act on Slater determinants and not on single-particle
states. If we have $N$ particles and retain $M$ single-particle
states, this requires the solution of $C_N^M-1$ elements of the
state vector (with $C_N^M=M!/N!(M-N)!$ and the $-1$ comes from the
normalization condition). In addition, one has to average over an
amount, call it $m$, of different realizations of the stochastic
process.~\footnote{A density-matrix formalism would be even more
computationally demanding, requiring the solution of $(C_N^M
+2)\times (C_N^M-1)/2$ coupled differential equations, even after
taking into account the constraints of hermiticity and unit trace of
the density matrix.} The problem thus scales exponentially with the
number of particles.

However, it was recently suggested in Ref.~\onlinecite{Pershin2008}
that for operators of the type $\hat A=\sum_j \hat A_j$, sum over
single-particle operators (like, e.g., the density or current
density), the expectation value of $\hat A$ over a many-particle
non-interacting state with dissipation can be approximated as a sum
of single-particle expectation values of $\hat A_j$ over an ensemble
of $N$ single-particle systems with specific single-particle
dissipation operators. In particular, the agreement between the
exact many-body calculation and the approximate single-particle
scheme has been found to be excellent for the current
density.~\cite{Pershin2008} We refer the reader to
Ref.~\onlinecite{Pershin2008} for the numerical demonstration of
this scheme and its analytical justification. The physical reason
behind it is that, due to the coupling between the system and the
environment, highly-correlated states are unlikely to form.

Here, for numerical convenience, we will adopt the same {\em ansatz}
which in the present case reads,
\begin{equation}
\overline{\langle \Psi_{KS}|\hat A| \Psi_{KS}\rangle} \simeq
\sum_{j=1}^N\overline{\langle \phi^j_{KS}|\hat A_j|
\phi^j_{KS}\rangle},\label{ansatzsp}
\end{equation}
with $|\phi^j_{KS}\rangle$ single-particle KS states solutions of
\begin{eqnarray}
d|\phi^j_{KS}\rangle &=&\left[-i \frac{\left( \hat
p+eA_{eff}(r,t)\right)^2}{2m}  -\frac12 \hat V_{sp}^\dagger \hat
V_{sp}\right]|\phi^j_{KS}\rangle dt\nonumber \\
& &+\hat V_{sp} |\phi^j_{KS}\rangle dW(t),
\label{kssesp}
\end{eqnarray}
with $\hat V_{sp}$ an operator acting on single particle states (see
Refs.~\onlinecite{Pershin2008} and~\onlinecite{Bushong2007} and next
section for explicit examples of such operator).~\footnote{Note that
the theorem of S-TDCDFT is still valid, and Eq.~(\ref{ansatzsp})
would be exact (and not an approximation), if we choose the bath
operators to act on single-particle states (or the density) to begin
with.}

For convenience, also in the present case we can normalize the
single-particle KS states for every realization of the stochastic
process by defining
\begin{equation}
|\tilde{\phi}^j_{KS}\rangle
=\frac{|\phi^j_{KS}\rangle}{||\phi^j_{KS}||}
\end{equation}
and thus solve the non-linear SSE
\begin{eqnarray}
d|\tilde{\phi}^j_{KS}\rangle &=&\left[-i \frac{\left( \hat
p+eA_{eff}(r,t)\right)^2}{2m}  -\frac12 \hat V_{sp}^\dagger \hat
V_{sp}\right . \nonumber \\
& & \left . +R_j\hat V_{sp} -\frac12 R_j^2\hat 1\right]|
\tilde{\phi}^j_{KS}
\rangle dt\nonumber \\
& &+(\hat V_{sp} - R_j\hat 1)|\tilde{\phi}^j_{KS}\rangle dW(t),
\label{kssespnl}
\end{eqnarray}
where
\begin{equation}
R_j=\frac{1}{2}\frac{\langle \phi^j_{KS}|(\hat V_{sp}^{\dag}+\hat
V_{sp})|\phi^j_{KS}\rangle}{||\phi^j_{KS}||^2}.
\end{equation}

The discretization of these equations is then done similarly to what
we have explained in the previous section.

\section{An example: A gas of linear harmonic oscillators}
\label{bose}

Stochastic-TDCDFT has been applied to the study of decay of exited
He atoms and its connection to quantum-measurement
theory.\cite{Bushong2007} It can describe the dynamics of bosons as
well. In this section, we apply it to the analysis of the dynamics
of an interacting 1D Bose gas confined in a harmonic potential and
coupled to a uniform external environment that forces the gas
towards some steady state. Since neither the external potential nor
the bath are time-dependent, we expect that the boson gas reaches a
steady state configuration when coupled with the uniform external
bath. Finally, we assume that the bath forces the system towards
certain eigenstates of the instantaneous interacting boson
Hamiltonian. The bosons are interacting via a two-body contact
potential, i.e., $U_{int}(x,x')\propto \delta(x-x')$. This potential
correctly describes the important case of Alkali gases in which the
Bose-Einstein condensation has been experimentally observed.
\cite{Davis1995,Anderson1995,Dalfovo1999}

The purpose of this section is to compare the dynamics of the boson gas
obtained from the SSE [Eq.(\ref{stochasticse-fin})] and the quantum
master equation [Eq. (\ref{density-matrix})]. For this reason the value
of the physical parameters (the strengths of the confining potential,
the particle-particle interaction, and the system-bath coupling) is
arbitrary and chosen only for the sake of this comparison.
We will report elsewhere a more realistic study of the dynamics
of this important physical system.
\footnote{In calculating
the time evolution with the SSE we make use of the techniques
discussed in Sec.~\ref{numeric}.} When no interaction between
particles is included both approaches are clearly equivalent.
However, when interactions are included, the Hamiltonian of the
system becomes stochastic and, as previously discussed, the quantum
master equation does not take into account correctly the statistical
correlations induced by the bath, while the SSE naturally accounts
for the the stochasticity introduced in the Hamiltonian by the
interaction potential. In fact, we find that when both the initial
and final state are pure, both approaches provide the same
equilibrium state. However, the corresponding {\em dynamics} are
different. In particular, the relaxation time obtained from the
evolution of the density matrix is shorter than the relaxation time
obtained from the average over many realizations of the dynamics
obtained from the SSE.

The differences between the two approaches are even more striking
when we consider the evolution towards a state that contains at
least two major contributions coming from different states. In this
case, also the final steady states obtained from the density matrix
and the SSE are different.

These cases exemplify what we have discussed all along: if one
insists on using a closed equation of motion for the density matrix
of the type~(\ref{density-matrix}) with stochastic Hamiltonians,
uncontrolled approximations are introduced which lead to an
incorrect dynamics.

\subsection{Macroscopic occupation of the ground state}
We begin with the study of the dynamics of the macroscopic
occupation of the ground state induced by energy dissipation towards
the degrees of freedom of an external bath. The external bath forces
the system to reach a state of zero temperature or minimal free
energy, i.e., the ground state of the Hamiltonian. One possible form
of this bath operator is, in a basis set that makes the Hamiltonian
diagonal at each instance of time,~\cite{Bushong2007}
\begin{equation}
\hat V\equiv\delta \left(\begin{array}{ccccc}
0 & 1 & 1 & 1 & \ldots \\
0 & 0 & 0 & 0 & \ldots \\
\vdots & \vdots & \vdots & \vdots & \vdots \\
0 & 0 & 0 & 0 & \ldots
\end{array}
\right),
\label{bath_ground}
\end{equation}
where $\delta$ is a coupling constant with dimensions of the square
root of a frequency [we set $\delta=\sqrt{\omega_0}$ in what
follows, with $\omega_0$ the frequency of the harmonic confining
potential -- see Eq.~(\ref{harmonic})]. We do not expect that this
operator fulfills Eq.(\ref{bathcondition}) since in a real-space
representation it would allow for the localization of the particles
without an effective current between two distinct points in space.
In this section, however, we are more interested in the kind of
dynamics this operator generates in our quantum system, and the
comparison with the dynamics obtained from the quantum master
equation. We expect, indeed, that the condition
Eq.(\ref{bathcondition}) is violated both in the SSE and in the
quantum master dynamics.

The operator (\ref{bath_ground}) mimics the energy dissipation in
the system, with the external bath absorbing the bosons' excess
energy and cooling down the boson gas. One could argue that this is
the generalization to the many-state system of the bath considered
in Ref.~(\onlinecite{vanKampen}). We can thus conclude that the
effective temperature of the bath we consider here is zero.

The Hamiltonian of the boson system (in second quantization) when the
bath is not present reads
\begin{eqnarray}
\hat H&=&\int
dx~\psi^\dagger(x)\left(-\frac{1}{2m}\frac{d^2}{dx^2}+V_{ext}(x)\right)
\psi(x)\nonumber\\
&&+\int dxdx'~ \psi^\dagger(x)\psi^\dagger(x')
U_{int}(x-x')\psi(x')\psi(x),
\end{eqnarray}
where $\psi(x)$ destroys a boson at position $x$, $V_{ext}(x)$ is a
confining potential, and $U_{int}(x-x')$ is the boson-boson
interaction potential. For dilute boson atomic gases the interaction
potential can be substituted with the contact potential, i.e.,
\begin{equation}
U_{int}(x-x')=g_0(N-1)\delta(x-x')=\tilde{g}\delta(x-x')\label{Hint}
\end{equation}
where $g_0$ is determined by the scattering length of the
boson-boson collision in the dilute gas, and $N$ is the total number
of bosons in the trap, so that $||\psi||$=1.\cite{Dalfovo1999}

With standard techniques, and in the Hartree approximation, we can go
from the equation of motion for the annihilation operators to the equation of
motion for the state of the system $\Psi(x,t)$, when the external
bath is not coupled to the boson gas,
\begin{equation}
i\partial_t
\Psi(x,t)=\left[-\frac{1}{2m}\frac{d^2}{dx^2}+V_{ext}(x)\right]\Psi(x,t)
+\tilde{g} n(x,t)\Psi(x,t)
\label{GP}
\end{equation}
where $n(x,t)=|\Psi(x,t)|^2$ is the single-particle density of the
boson gas.\cite{Gross1963,Ginzburg1958}

Equation~(\ref{GP}) (and its generalization to 2 and 3 dimensions)
has received a lot of attention since it correctly describes the
dynamics of a Bose-Einstein
condensate.\cite{Davis1995,Anderson1995,Dalfovo1999}

In the following we will focus on the case of a 1D harmonic confining
potential, i.e.,
\begin{equation}
V_{ext}(x)=\frac12 m\omega_0^2 x^2.\label{harmonic}
\end{equation}
A harmonic confinement is created, e.g., in the magneto-optical
traps used in the experimental realization of the Bose-Einstein
condensation of dilute boson Alkali
gases.\cite{Davis1995,Anderson1995,Dalfovo1999}

When the boson system is coupled to the external environment, we assume
that the Hamiltonian is not affected by the coupling and the state of the system $\Psi(x,t)$,
that is now stochastic,
evolves according to the SSE
\begin{widetext}
\begin{equation}
d\Psi(x,t)=-i\left(-\frac{1}{2m}\frac{d^2}{dx^2}+\frac12 m\omega_0^2 x^2
+g n(x,t)\right)\Psi(x,t)dt-\frac12 \hat V^\dagger \hat V\Psi(x,t)dt
+\hat V\Psi(x,t)dW
\label{sse-gp_notnorm}
\end{equation}
where this equation of motion has to be interpreted in accordance
to the discussion of the previous Sections. For numerical
convenience, we rescale this equation in terms of the physical quantities
$\omega_0$, $x_0=1/\sqrt{m\omega_0}$, and $g=\tilde{g}/x_0$ to arrive at
\begin{equation}
d\Psi(x,t)=-i\omega_0\left(-\frac{x_0^2}{2}\frac{d^2}{dx^2}+\frac{x^2}{2 x_0^2}
+\frac{g}{\omega_0} n(x,t)x_0\right)\Psi(x,t)dt-\frac12 \hat V^\dagger \hat V\Psi(x,t)dt
+\hat V\Psi(x,t)dW.
\label{sse-gp}
\end{equation}
\end{widetext}

We begin by considering the case of non-interacting bosons, i.e., we
set $g = 0$. In this case, the Hamiltonian admits a natural complete
basis, the set of Hermite-Gauss wave-functions
\begin{equation}
\varphi_j (x)=\frac{1}{\sqrt{x_0 2^j j! \sqrt{\pi}}}H_{j}(x/x_0)
e^{-x^2/2x_0^2}
\end{equation}
where the polynomials $\{H_j\}$ satisfy the recursion relation
\begin{equation}
H_{j+1}(x)=2 x H_j(x)-H_{j-1}
\end{equation}
and $H_0(x)=1$, $H_1(x)=2x$. If we expand the wave-function
$\Psi(x,t)=\sum_{j}a_j(t)\varphi_j(x)$, and make use of the
orthonormality properties of the Hermite-Gauss wave-functions, we
obtain the (stochastic) dynamical equation for the coefficients
$a_j$,
\begin{equation}
d a_i=\sum_j \left(H_{ij}a_j+\frac12 (\hat V^\dagger \hat
V)_{ij}a_j\right) dt+ dW \sum_j \hat V_{ij}a_j
\label{freea}
\end{equation}
where $H_{ij}=(j+1/2)\omega_0 \delta_{ij}$ and $\hat V$ is given by
Eq.~(\ref{bath_ground}).

Together with Eq.~(\ref{freea}) we can study the dynamics of the
density matrix via the quantum master equation
Eq.~(\ref{density-matrix}), which in the same spatial representation
as Eq.~(\ref{freea}) reads
\begin{widetext}
\begin{equation}
\partial_t \rho_{ij}=-i\sum_k
\left(H_{ik}\rho_{kj}-\rho_{ik}H_{kj}\right)+\sum_{k,k'}\left(\hat
V_{ik}\rho_{kk'}\hat V^\dagger_{k'j}-\frac12 \hat V^\dagger_{ik}\hat
V_{kk'}\rho_{k'j}-\frac12 \rho_{ik}\hat V^\dagger_{kk'}\hat
V_{k'j}\right), \label{freerho}
\end{equation}
\end{widetext}
The connection between Eq.~(\ref{freea}) and Eq.~(\ref{freerho}) is
established by the identity $\rho_{ij}=a^*_i a_j$ valid for any pair
of indexes $i$ and $j$. We solve Eq.~(\ref{freea}) numerically with
a 4th order Runge-Kutta evolution scheme, after we have mapped the
dynamics to its norm-preserving equivalent form (see
Sec.~\ref{numeric} and the discussion
therein).\cite{Ghirardi1990,Breuer2002} For consistency we solve the
master equation with a 2nd order Runge-Kutta evolution scheme (in
fact with the more refined Heun's scheme).\cite{Press1992}

\begin{figure}[ht!]
\includegraphics[clip,width=8cm]{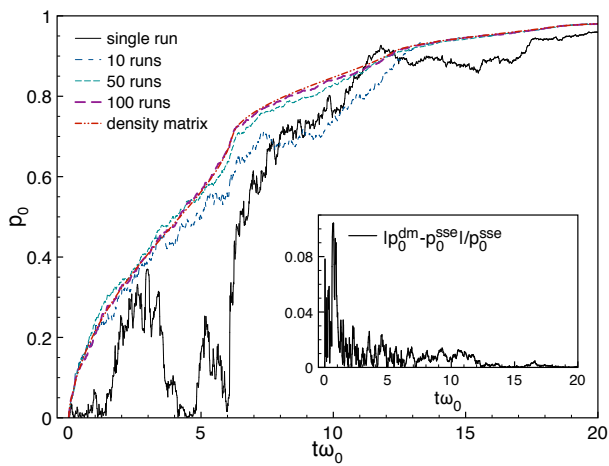}
\caption{(Color online) {\it Non-interacting bosons --} Occupation
probability of the ground state versus time calculated from the
evolution of the state via the SSE averaged over different runs (1,
10, 50 and 100) and via the equation of motion for the density
matrix for the non-interacting boson case. It is evident that with
only 50 realizations the agreement between the SSE and the density
matrix equation is excellent. In the inset we show the relative
difference between the two dynamics for 100 realizations of the
stochastic process.} \label{ground_occupation}
\end{figure}

We report in Fig.~(\ref{ground_occupation}) the dynamics of the
probability of occupation of the ground state, $p_0=|a_0|^2=\rho_{00}$ for various
realizations of the stochastic field in Eq.~(\ref{freea}) together with the dynamics
obtained from the evolution of the density matrix (\ref{freerho}).
Here, we have included the first 20 levels of the free Hamiltonian, and we have chosen as
initial condition
$a_{20}(0)=1$ and set the other coefficients to zero.
We have set the mass of the particles to $1$, and used a time step
$\omega_0 \Delta t=20/2^{15}=6\times 10^{-4}$. A further decrease of
this time step does not affect the results significantly. From
Fig.~\ref{ground_occupation}, it is evident that when we collect a
large enough statistics the results of the SSE and the master
equation coincide for the non-interacting boson case: Already for 50
runs of the SSE the difference between the two dynamics almost
vanishes.\footnote{We expect that, for non-interacting particles,
the deviation between the dynamics obtained via the density matrix
equation and the SSE, scales as $1/\sqrt{m}$ if $m$ is the number of
independent runs on which we average the SSE.} In the inset of
Fig.~\ref{ground_occupation} we report the relative difference
between the occupation number of the ground state with the two
dynamics, $|p_0^{dm}-p_0^{sse}|/p_0^{sse}$. We see that this
difference, for 100 runs, is generally lower than $5\%$, a quite
satisfactory result.

\begin{figure}[ht!]
\includegraphics[clip,width=8cm]{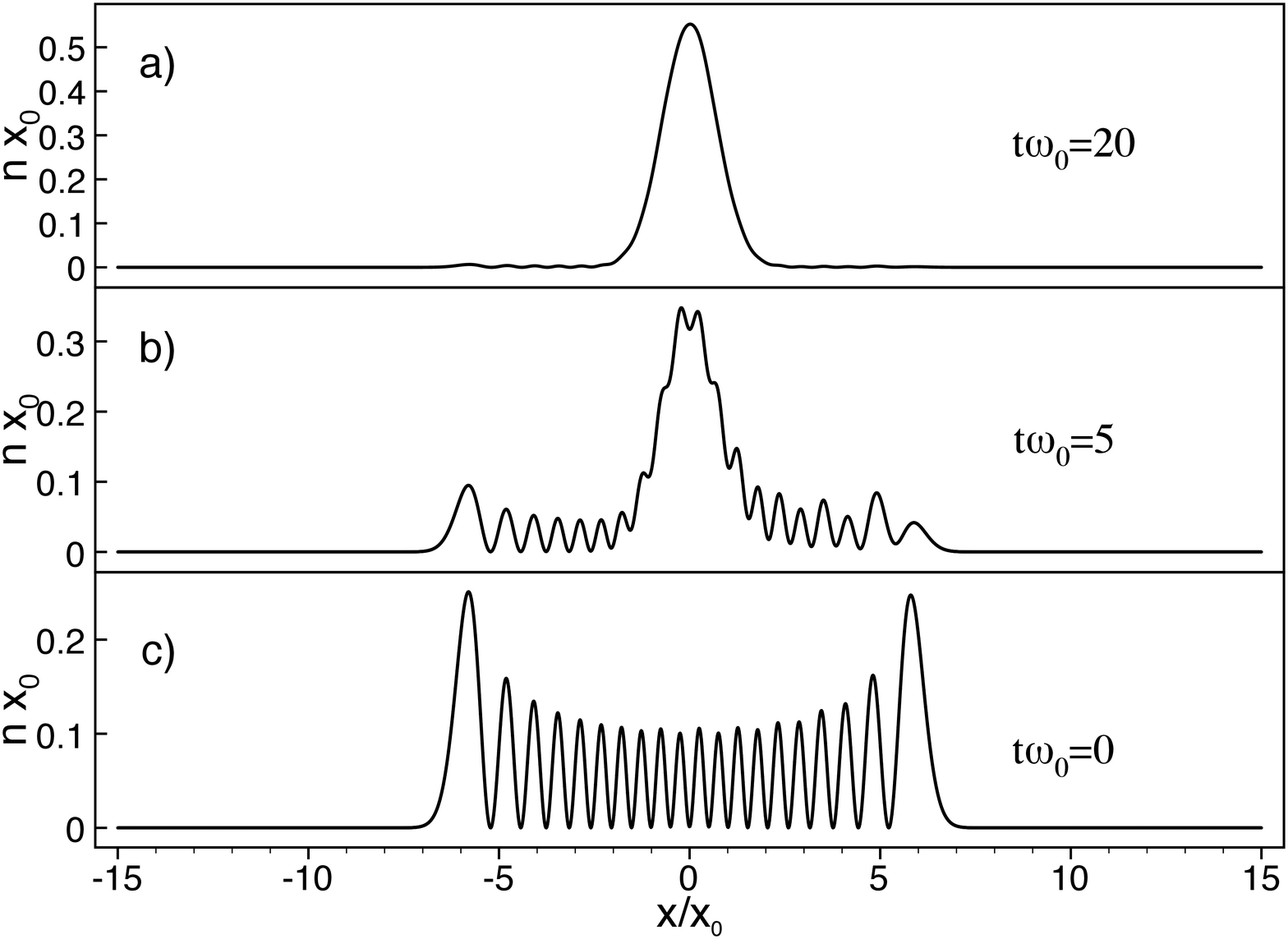}
\caption{{\it Non-interacting bosons --} Plot of the averaged
density profile, $n(x)\times x_0^2$,
for various instances of time calculated from the
SSE. The system evolves from the maximum occupation of the highest
excited state (panel c), to the maximum occupation of the ground
state. We have averaged over 100 realizations of the stochastic
process.} \label{density_plot}
\end{figure}

In Fig.~\ref{density_plot}, we report the density profile for the
system at different instances of time obtained from the SSE.
Starting from a pure state, where the highest energy state is
occupied (panel $c$), the system relaxes towards the ground state.
As it is clear from panel $a)$ in Fig.~\ref{density_plot} the
system, at $t\omega_0=20$, still occupies certain high energy states
[see, e.g., the tail at $x<0$ of panel $a)$].

We now turn on the particle-particle interaction $U_{int}$. This
corresponds to adding to the free Hamiltonian $H_{i,j}$ an
interaction part, $H^{int}$ that in the basis representation of the Gauss-Hermite
polynomials reads
\begin{equation}
H^{int}_{i,j}=\sum_{k,q}F_{i,j;k,q} a_k^* a_q
\end{equation}
where $F_{i,j;k,q}$ is the 4th-rank tensor defined as
\begin{equation}
F_{i,j;k,q}=g\int_{-\infty}^\infty dx~ H_i(x)H_j(x)H_k(x)H_q(x)e^{-2x^2}.
\end{equation}
A long but straightforward calculation brings us to an explicit
expression of $F_{i,j;k,q}$ in terms of Euler gamma functions and a
hypergeometric function.\cite{Lord1949,Abramowitz1964} It can be
shown that the hypergeometric function reduces to the summation of a
few -- at most $\min(i,j)$ -- terms. In the case of the density
matrix approach the interaction Hamiltonian is immediately written
as
\begin{equation}
H^{int}_{i,j}=\sum_{k,q}F_{i,j;k,q}\rho_{k,q}.
\end{equation}

In solving the dynamics of the system described either by the
SSE~(\ref{sse-gp}) or the master equation~(\ref{freerho}), we have
assumed that at any instance of time the bath operator brings the
system towards the instantaneous ground state of the {\it
interacting} Hamiltonian $H_{i,j}+H_{i,j}^{int}$.

\begin{figure}[ht!]
\includegraphics[clip,width=8.5cm]{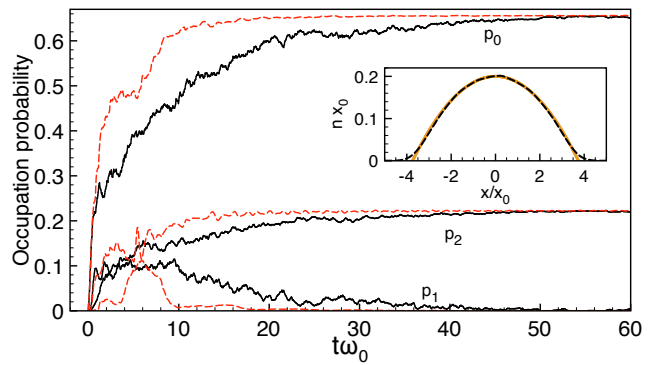}
\caption{(Color online) {\it Interacting bosons --} Occupation
probability of the first 3 lowest energy levels of the
non-interacting Hamiltonian versus time calculated via the SSE
(black, solid lines) averaged over 100 independent runs, and via the
equation of motion for the density matrix (red, dashed lines). The
time it takes the system to reach steady state is different for the
density matrix approach and the SSE, with the former underestimating
the relaxation time. This is due to the inclusion in the master
equation of the average density in the interaction potential, thus
neglecting important fluctuations that can slow down the relaxation
dynamics. In the inset we compare the equilibrium density (black,
dashed line) with the one obtained from the Thomas-Fermi
approximation to the ground state (orange, solid line).}
\label{projection}
\end{figure}

In Fig.~\ref{projection} we plot the occupation probability $p_j(t)$
of the state $j$ for the first 3 levels of the free Hamiltonian
($p_j(t)=|a_j(t)|^2$ from the SSE or $p_j(t)=\rho_{j,j}(t)$ from the
density matrix). We have assumed an interaction of strength
$g/\omega_0=5$, and a time step $\omega_0 \Delta t=60/2^{17}$ and we
have performed 100 independent runs of the SSE. While it evident
that the system reaches the same steady state according to the two
equations,~\footnote{The initial state is pure and the bath is
selecting only a particular state thus forcing the system towards
another pure state. Moreover, we can prove that if the system
evolves from the ground state, the stochastic part vanishes on this
state, and then the boson gas remains in the ground state of the
interacting Hamiltonian.} it is also clear that the state calculated
with the SSE relaxes slower than the state obtained from the density
matrix equation. This is a spurious effect in the density matrix
dynamics where the average density defines the interaction
potential. This does not take into account the fluctuations of the
state, and hence of the stochastic Hamiltonian.

We have also tested that the steady state reached during the
dynamics is consistent with the theory of the eigenstates of the
Gross-Pitaevskii equation. \cite{DAgosta2000,Dalfovo1999} In
particular, the ground state of the interacting system, when the
interaction is strong, can be obtained by neglecting the kinetic
contribution to the Hamiltonian. In this case, a good approximation
to the ground state density reads
\begin{eqnarray}
|\Psi_0(x)|^2&=&\frac{\mu-1/2 m\omega_0^2 x^2}{gx_0}\theta \left(\mu-1/2 m\omega_0^2 x^2\right)\nonumber \\
&&+\mathrm{terms~proportional~to }~1/g^2
\label{thomas-fermi}
\end{eqnarray}
where $\mu$, the chemical potential, is determined by the
normalization condition, and $\theta(x)=0$ if $x<0$ and
$\theta(x)=1$ if $x>0$.

In the inset of Fig.~\ref{projection} we plot the density obtained
at $t\omega_0=60$ from the SSE (black, dashed line) and the density
obtained from the approximation (\ref{thomas-fermi}) (orange, solid
line). Notice that the value of the parameters $g$ and $\mu$ have
been obtained from the best fit with the numerics: indeed one can
show that the approximation (\ref{thomas-fermi}) is exact in the
limit of very large interaction,\cite{DAgosta2000,Dalfovo1999} which
is not reached in our calculations.

\begin{figure}[ht!]
\includegraphics[clip,width=8cm]{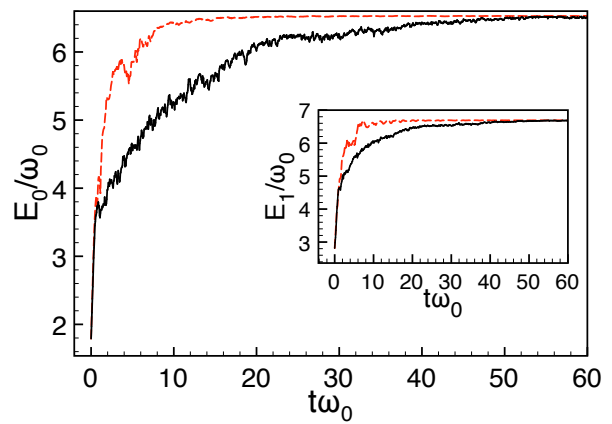}
\caption{(Color online) {\it Interacting bosons --} Time evolution
of the energy of the ground state of the Gross-Pitaevskii
Hamiltonian as calculated from the SSE (black, solid line) and
master equation (red, dashed line). The final value of the energy is
the same, but the relaxation dynamics is different in the two
formalisms with the master equation considerably underestimating the
relaxation time. In the inset we report the dynamics of the first
excited state of the Gross-Pitaevskii Hamiltonian obtained from the
SSE (black, solid line) and the master equation (red, dashed line).}
\label{energy}
\end{figure}

In Fig.~\ref{energy} we report the value of the ground state energy
of the interacting Hamiltonian versus time as calculated from the
SSE and the master equation. Again the difference between the
relaxation times calculated from the two dynamics is evident. In the
inset of Fig.~\ref{energy} we report the energy of the first excited
state.

To summarize this section, we have described the dynamics of the
relaxation of a confined 1D boson system towards the ground state
induced by a given external bath. The final state we have obtained
is consistent with the eigenstate of the 1D Gross-Pitaevskii
equation. Our main result is that, although the SSE and the master
equation reach the same final state, the {\it dynamics} described by
these equations show important differences, and physical quantities,
like, e.g., the relaxation time, differ. In particular, the density
matrix approach, which at any instant of time employs the average
density to construct the interaction Hamiltonian, underestimates the
fluctuations induced by the bath on the {\it stochastic}
Hamiltonian. These fluctuations are correctly taken into account in
the SSE.

\subsection{Competition between states}
\begin{figure}[ht!]
\includegraphics[clip,width=8.5cm]{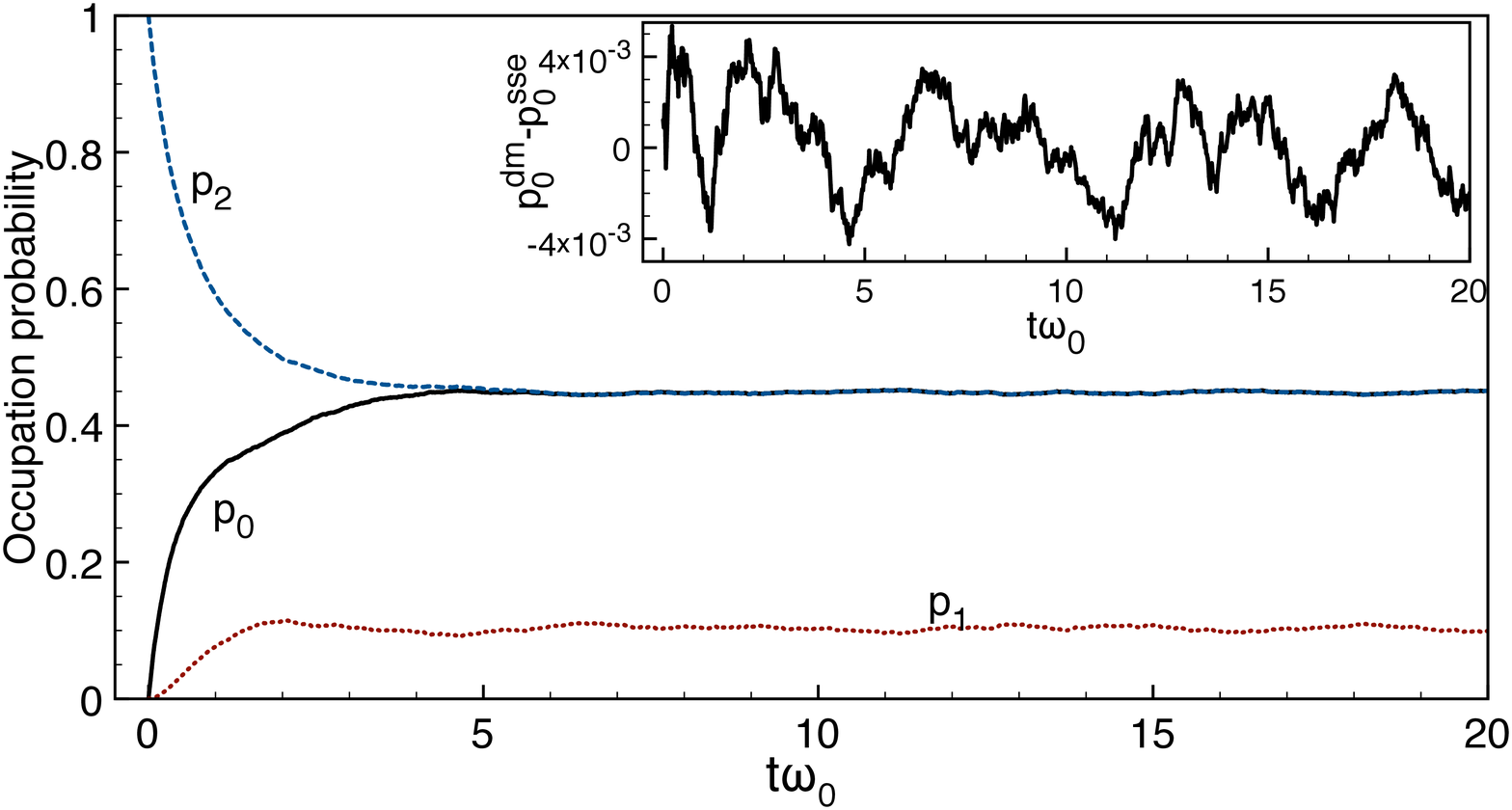}
\caption{(Color online) {\it Non-interacting bosons --} The
occupation probabilities for a three level system calculated via the
SSE~(\ref{sse-gp}) for non-interacting bosons ($g=0$). The results
obtained via the master equation for the density matrix are
indistinguishable on this scale from those obtained with the SSE. In
the inset we show the difference between the occupation
probabilities of the lowest energy level calculated from the SSE and
the master equation. This difference is in modulus lower than
$5\times 10^{-3}$ at any instant of time.} \label{3level-free}
\end{figure}
Let us now consider the more common case in which the environment
drives the system toward a mixed steady state. To simplify the
discussion we consider only three single-particle levels and the
bath operator forces the system towards two different states. We
choose, in a basis in which the Hamiltonian is diagonal, the
operator
\begin{equation}
V=\delta \left(\begin{array}{ccccc}
0 & 1 & 1 \\
0 & 0 & 0 \\
1 & 1 & 0
\end{array}
\right),
\label{bath_compete}
\end{equation}
i.e., the operator drives the system, with equal strength, towards
the lowest and highest energy levels of the interacting Hamiltonian.
As we will see, the final state is a superposition of these two
states with a significant contribution coming from the middle level.
At first glance this might seem surprising. However, we have to
remember that, e.g., in the quantum master equation, the equilibrium
states are determined by the kernel of the super-operator. This
super-operator contains powers of the operator $V$, that in turn
contains a finite contribution from the middle level. A similar
reasoning applies to the SSE.

To begin with our analysis of this system, we consider the
non-interacting case $g=0$; we set as before
$\delta=\sqrt{\omega_0}$; and we start from the fully occupied
highest energy level, i.e., $a_3(0)=1$. In Fig.~\ref{3level-free} we
plot the occupation probabilities for the 3 levels calculated via
the SSE~(\ref{sse-gp}). In this case, to reduce the stochastic noise
even further, we have performed 1000 independent runs of the SSE and
used, in both dynamics, $\omega_0\Delta t=20/2^{14}$. As we can see
from Fig.~\ref{3level-free}, at steady state the bath operator
forces the system to occupy the lowest and the highest energy levels
with equal probability, while a finite occupation probability of the
middle level appears. This mixing prevents the system to reach a
pure steady state and some finite correlation between the energy
levels, that appears for example in the finite off-diagonal elements
of the density matrix, persists in the long-time regime.

Again, for this non-interacting case the dynamics obtained from the
SSE and master equation are indistinguishable on the scale of the
plot of Fig.~\ref{3level-free}.\footnote{Only the dynamics obtained
from the SSE is reported in Fig.~\ref{3level-free}.} In the inset of
Fig.~(\ref{3level-free}), we report the difference between the
ground state occupation probability as calculated from the SSE and
from the density matrix approach. This difference is, in amplitude,
smaller than $5\times10^{-3}$, and by increasing the number of
independent runs, it decreases. To test our numerical code, we have
also compared the numerical solution with the exact dynamics
obtained from the analytical solution of the master equation (which
is feasible because we have only three states). Since the numerical
and analytical solutions are essentially the same, we do not find
necessary to report the analytical solution here.
\begin{figure}[t!]
\includegraphics[width=8.5cm,clip]{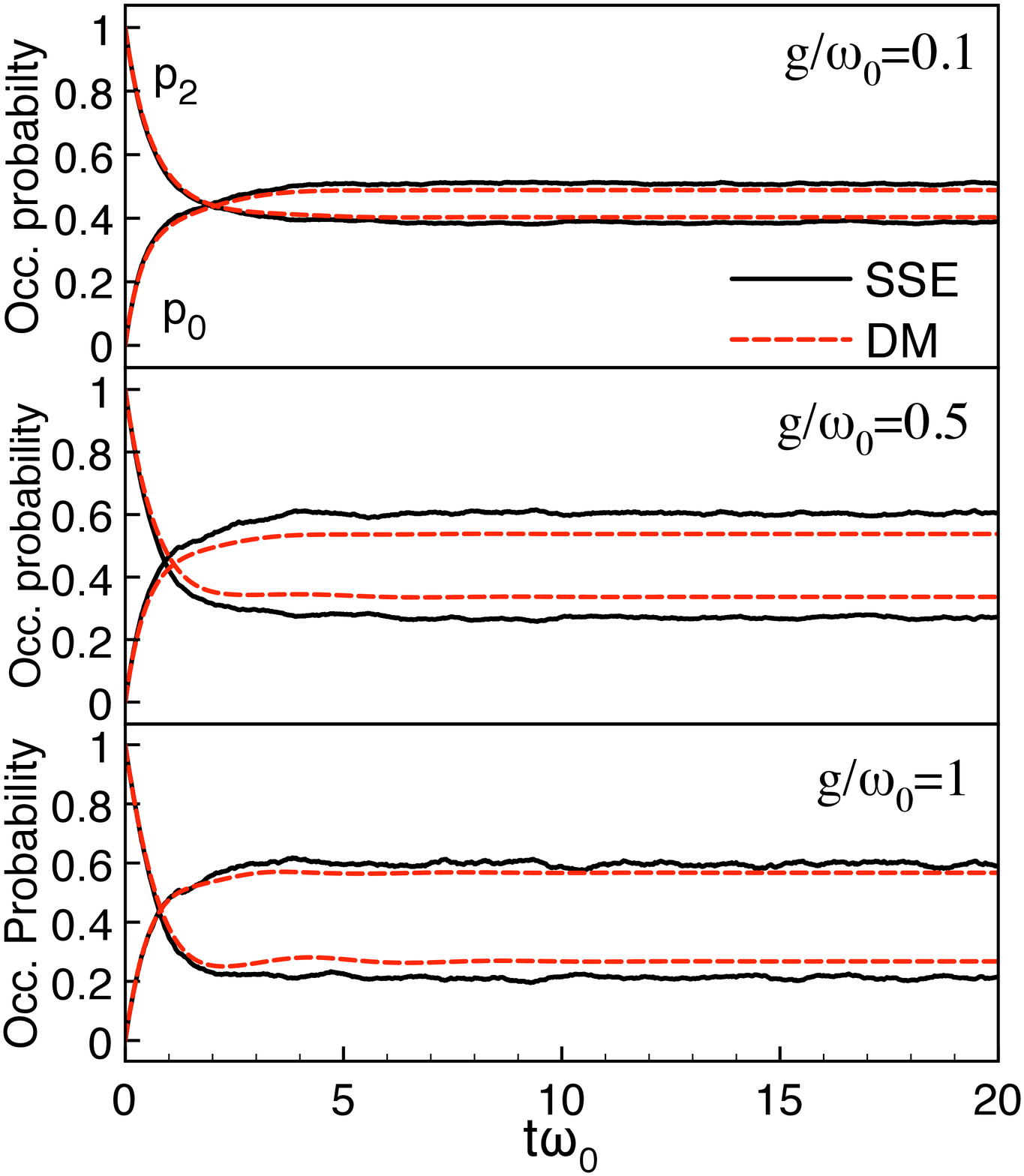}
\caption{(Color online) {\it Interacting bosons --} Plot of the
dynamics of the occupation numbers of the lowest and highest energy
level, $p_0$ and $p_2$, respectively, calculated from the
SSE~(\ref{sse-gp}) (black, solid line) and the master equation for
the density matrix~(\ref{density-matrix}) (red, dashed line). Panel
a), $g/\omega_0=0.1$: for small interaction the two dynamics show
small differences. Panel b), $g\omega_0=0.5$: for intermediate
interaction strength the differences between the two dynamics is a
large fraction of the occupation number. Panel c), for large
interaction $g/\omega_0=1$ the difference between the two dynamics
for the lower level decreases. In this particular case, this is due
to the presence of the second energy level (not shown in the figure)
that is little affected by the interaction. } \label{3int}
\end{figure}

We now turn on the particle-particle interaction~(\ref{Hint}).
Fig.~\ref{3int} reports the time evolution of the occupation number
of the lowest and highest energy levels of the free Hamiltonian for
different strengths of the particle-particle interaction. As
expected, the interaction opens a gap in the occupation numbers
between the highest and lowest energy levels. Most importantly, we
see that for intermediate values of the interaction the steady
states calculated with the SSE and the master equation differ. This
difference is not monotonic with the interaction, and it is
state-dependent. We see indeed that for relatively strong
interaction $g/\omega_0\ge 1$, this difference is smaller than for
$g/\omega_0=0.5$; more so for the lowest state than the highest one.
This is due to the fact that the middle energy level (not shown in
the figure), which is almost unaffected by the variation of the
interaction strength and whose dynamics is almost the same for the
SSE and the master equation, ``blocks'' the transformation of the
highest energy level to low occupation numbers. For very strong
interaction $g/\omega_0=5$ (not shown in the figure), the occupation
numbers calculates via the SSE and the density matrix approach,
almost coincide.

The above example shows that when the bath drives the system towards
a mixed state, also the \emph{final states} (not just the dynamics)
obtained from the density matrix according to the master
equation~(\ref{density-matrix}) and the SSE may be \emph{different}.
In the particular case considered here, this is due to the fact that
the final state is sensitive to the frequency of the confining
potential (as can be shown with the exact analytical solution of the
non-interacting system). The SSE and the master equation create
different effective interaction potentials that renormalize the
frequency of the confining harmonic potential. This different
renormalization shows up in the different steady states. This
important difference is again due to the fact that in the master
equation the interactions are included using the average particle
density, thus neglecting the true stochasticity of the Hamiltonian.
Small differences in the effective potential (confining plus
interaction) thus results in macroscopic differences in the steady
states. The fact that the dynamics of the interacting system
described by the master equation~(\ref{density-matrix}) is so
sensitive to the interaction potential and does not reproduce
correctly the dynamics and/or the steady states of the system
undermines the applicability of an equation of motion for the
density matrix to the stochastic extension of TDDFT and TDCDFT.

\section{Conclusions}\label{conclude}
In this paper, we have discussed in detail a functional theory of
open quantum systems we have named stochastic TDCDFT. This theory,
based on a theorem we have previously proved in
Ref.~\onlinecite{Diventra2007}, extends DFT to the dynamical
interaction of quantum systems open to external environments, when
the latter satisfy a memory-less dynamics. The starting point of the
theory is a stochastic Schr\"odinger equation for the $N-$particle
state vector, which provides a conceptually transparent way of
describing open quantum systems.

We have discussed the mathematical assumptions of the theory, the
numerical solution of the corresponding equations of motion, and
compared it to a possible formulation in terms of a density-matrix
approach based on quantum master equations. We have shown that due
to the dependence of the KS Hamiltonian on microscopic degrees of
freedom, and its time-dependence, a density-matrix approach to a
stochastic DFT is {\em not} a solid alternative to this problem. In
fact, due to these conditions, there is not necessarily a closed
equation of motion for the density matrix, and if one insists on
using a quantum master equation, the solutions of such an equation
may not be physical for all cases.

As an example of application, we have used the theory to study the
dynamics of a 1D gas of excited bosons confined in a harmonic
potential and in contact with an external bath. This is a problem
previously inaccessible by standard DFT. Along similar lines, we
expect this theory to find application in a wide range of problems
where DFT methods could not be applied, such as energy transport and
dissipation, dephasing induced by an environment, quantum
measurement and quantum information theory, phase transitions driven
by dissipative effects, etc.

From here, an interesting (and non-trivial) extension of stochastic
TDCDFT would be to environments with finite auto-correlation times.
This leads to non-Markovian dynamics with memory kernels and more
complex stochastic Schr\"odinger
equations.~\cite{Gaspard1999,Maniscalco2004} If a similar theorem as
that we have demonstrated here can be proved for these cases as
well, we could study an even larger class of open quantum system
problems, where memory effects in the bath dynamics are of
particular importance.

Another possible extension of the theory would be to investigate the
noise properties of the quantum system. This would provide even more
information on the system dynamics. An extension of S-TDCDFT to this
problem seems possible but not trivial. The reason is because the
noise is an $n$-time correlation function (where $n$ indicates the
moments of the observable), and as such it cannot be written simply
in terms of the expectation value of an observable. It is thus not
obvious what is the physical variable conjugated to the noise of
given moment. One could clearly calculate the moments of the current
using the present form of S-TDCDFT. How good this approximation is
compared to the exact noise (even if one knows the exact functional
of S-TDCDFT) is an issue that, like other applications of DFT beyond
its basic theorems (e.g., the assignment of a physical meaning to
the KS states), must be addressed at an ``empirical'' level by
comparing with experiments or available analytical results.

Finally, another important direction of study would be the
development of functionals in the presence of baths. Clearly, this
cannot be done for arbitrary baths, and specific cases, such as a
bath of harmonic oscillators, would be a good starting point. It
would be interesting to know if an approximate functional with a
clear physical interpretation can be obtained, and how different it
is from the functionals in the absence of bath interaction. Until
then, the best we can do is to apply the available functionals,
justify their use on the basis of the weak interaction between the
system and the environment, and compare the results with available
experimental data or analytical results.

\begin{acknowledgments}
We thank N. Bushong, Y. Pershin, Y. Dubi and G. Vignale for useful
discussions. This work has been supported by the Department of
Energy grant DE-FG02-05ER46204.
\end{acknowledgments}

\bibliography{mine,articles,books}
\end{document}